\documentclass[%
 reprint,
 amsmath,amssymb,
 aps,
floatfix,
]{revtex4-1}

\usepackage{graphicx}
\usepackage{dcolumn}
\usepackage{bm}

\makeatletter
\newcommand{\thickhline}{%
    \noalign {\ifnum 0=`}\fi \hrule height 1pt
    \futurelet \reserved@a \@xhline
}
\newcolumntype{"}{@{\hskip\tabcolsep\vrule width 1pt\hskip\tabcolsep}}
\makeatother

\begin{document}

\title{Limits on Magnetically Induced Faraday Rotation from Polarized $^3$He Atoms}

\author{J. Abney}
\author{M. Broering}
\author{M. Roy}
\author{W. Korsch}

\affiliation{%
 University of Kentucky
}%

\begin{abstract}
Faraday rotation has become a powerful tool in a large variety of physics applications. Most prominently, Faraday rotation can be used in precision magnetometry. Here we report measurements of gyromagnetic Faraday rotation on a dense, hyperpolarized $^3$He gas target. Theoretical calculations predict the rotations of linearly polarized light due to the magnetization of spin-1/2 particles are on the scale of 10$^{-7}$ radians. To maximize the signal, a $^3$He target designed to use with a multipass cavity is combined with a sensitive apparatus for polarimetry that can detect optical rotations on the order of 10$^{-8}$ radians. Although the expected results are well above the sensitivity for the given experimental conditions, no nuclear-spin induced rotation was observed. 
\end{abstract}

\maketitle


\section{\label{sec:level1}Introduction}

The Faraday effect is a well known dispersive phenomenon in which the plane of polarization of linearly polarized light is rotated while traversing a material placed in a magnetic field which is directed along the path of light propagation. Especially in recent years this circularly birefringent effect has evolved into a precision tool for a variety of applications. For example, it is used to measure permanent electric dipole moments in atoms \cite{FR_HgEDM}, to determine magnetic field strengths and directions in the intergalactic medium \cite{FR_ISM}, as part of precision magnetometers to test fundamental symmetries \cite{FR_sym_test}, to measure spin-noise \cite{FR_Spin_noise}, or to probe the process of light squeezing \cite{FR_spin_squeezing}, just to name a few. However, all these applications rely on the interaction of light with the electric charge and thus the rotations are induced by electric polarizabilities of the studied systems. Such effects are strongly enhanced when the probing light is monochromatic with a frequency that is slightly detuned from the center of an atomic resonance \cite{FR_resonant}. In addition to this electrically induced rotation a similar effect should exist due to the magnetic interaction of light \cite{Susan2008}. This so called gyromagnetic Faraday effect is significantly smaller and it exhibits a different frequency dependence than its electric counterpart. The experimental verification of this magnetic effect would open up new venues for studies of fundamental properties of Dark Matter, such as putting limits on possible anomalous magnetic moments of composite low-mass Dark Matter particles \cite{Susan2009}. Furthermore, it could be considered for the development of new ways to monitor the polarization of dense spin-polarized targets. In this paper we present measurements of the spin-induced Faraday rotation on a dense hyperpolarized $^3$He gas.

\section{The Gyromagnetic Faraday Effect}

The application of Faraday rotation appears to be exclusively devoted to the coupling of light to the electric polarizability of atoms and molecules. The theory of the interaction of light with electric free or bound charges is well established and the general behavior is described classically and quantum mechanically \cite{Buckingham}. However, the magnetic contribution to this effect has attracted less attention in the literature. It has been pointed out in several papers \cite{Susan2008, Russian_paper} that a rotation of the light can also be induced by magnetic moments. The only claimed observation has been in the infra-red on ferrite garnets \cite{Russian_paper,Russian_paper_2}. Such magnetic Faraday rotations are typically vastly smaller and they are experimentally more difficult to isolate. Ideally, in order to really investigate the magnetically induced rotation, a neutral particle with minimal intrinsic structure and an anomalous magnetic moment is the best candidate. Naturally free neutrons would be the perfect choice. However, the problem with neutrons is that they are unstable with a lifetime of about 880 s and therefore it is very difficult to produce dense targets. This is, for example, a major factor in the progress of improving the limit of the neutron permanent electric dipole moment \cite{nEDM_review,nEDM_review_2}. A system that has several of the advantages required here is helium-3. This helium isotope is predominantly in a singlet {\it {S}} ground state at room temperature, the single-electron excitation energy is relatively large ($\Delta E_{1S - 1P} \approx 21$ eV), and it has a spin-1/2 nuclear ground state with a magnetic moment of $- 2.12 \mu_N$. Additionally, two well developed technologies exist to produce relatively dense and nuclear spin-polarized targets with high values of polarization. Typically such targets are either polarized by means of metastability exchange optical pumping (MEOP) or spin exchange optical pumping (SEOP). State of the art polarization values of up to about 80\% have been reported at number densities reaching 7 amagats \cite{High_pol_He3}.

The theory for the gyromagnetic Faraday effect has been developed by Gardner and also Krinchik and Chetkin (see \cite{Susan2009, Russian_paper}). We summarize the main ideas in the following. As shown in \cite{Susan2009}, a Faraday rotation induced by a sample of magnetic moments is given by:

\begin{equation}
\begin{aligned}
\theta_{FR} = \frac{\mu_{0}}{\hbar c} \mu^2 \rho P L,
\end{aligned}
\label{eq:He_short_paper_1}
\end{equation}   

where $\mu$ denotes the magnetic moment of the particle, $\rho$ is number density of the sample, $P$ is the spin-polarization, and $L$ is the sample length. The quadratic dependence on the magnetic moment is due to the replacement of the magnetization of the target by its polarization, $M = \mu \rho P$. It should be emphasized that the induced rotation is, to leading order, independent of the frequency of the probing light. Meaning that the same optical rotation angle is expected for all wavelengths. This is in contrast to other magneto-optical rotations dependent which are dependent on the electric polarizability and typically have a $1/\lambda^2$ dependence. Optical rotations proportional to the polarization and having such a wavelength dependence have been observed before \cite{LiquidstateNMR,NSORxenon}, however such effects come from fundamentally different mechanisms than the gyromagnetic Faraday effect. Using typical numbers for existing polarized $^3$He targets, for example a density of 4.4 amagats, a target length of 40 cm, and a spin-polarization of 55\%, one would expect a spin induced rotation of about 120 nrad. In comparison, the electric Faraday effect from helium under the experimental conditions will be well below this at $\sim10^{-11}$ rad, which is an additional benefit to performing the experiment with helium. In order to measure the predictated small rotation angle, a dedicated experiment had to be conceived which combined such a polarized target with a highly sensitive Faraday rotation apparatus and a multipass cavity. 

\section{\label{sec:level2}$^3$He Target}

For the experiment, the method chosen to polarize $^3$He is spin exchange optical pumping. The $^3$He gas is contained inside an aluminosilicate glass cell of similar style to the SEOP hybrid cells used at Jefferson Lab for electron scattering experiments \cite{JLab_cell}. A spherical pumping chamber on top, which contains small amounts of rubidium and potassium besides the $^3$He, is connected by a transfer tube to a 39.45 cm long cylindrical target chamber on the bottom. Rb vapor in the pumping chamber is polarized with a high power, narrowband (120 GHz) 795 nm laser whose light is converted into circular polarization with quarter-wave plates mounted rotation stages. The Rb then transfers this polarization to K vapor via spin exchange collisions. Both alkali metals then polarize $^3$He with spin exchange collisions and the $^3$He gas slowly gains nuclear spin polarization in this manner. The pumping chamber is contained inside a forced air oven heated to 235$^{\circ}$C that is suspended on a platform which has vibration damping to reduce noise due to movements. Located below the oven is the target chamber which is in the path of the probe laser used to measure the Faraday rotation. Fig.(\ref{fig:He_short_paper_1}) shows the basic geometry of the cell. The target chamber is terminated by flat windows with a diameter of 2.54 cm on the end to ensure minimal divergence of the probe laser. These windows are designed for use with a multipass optical cavity so the probe laser can be passed through the chamber multiple times to increase the optical rotation angle. A static magnetic field, stabilized with a feedback system and running parallel to the length of the target chamber and the probe laser provides an axis to align the nuclear spins. An AC magnetic field transverse to the DC holding is used to flip the spins with adiabatic fast passage (AFP) frequency sweep NMR. The $^3$He polarization can then be measured with a pair of pick-up coils separated by a distance of 3.25 cm parallel to the target chamber. Absolute polarization is extracted by electron paramagnetic resonance (EPR) frequency shift where the small Zeeman shift in the alkali metal resonance frequencies due to the effective magnetic field of the polarized $^3$He is detected \cite{PhysRevA.58.3004}. Through EPR the polarization is measured in the pumping chamber and corrected for transport loss when comparing to the NMR measurement of the target chamber. The specific cell used for this experiment has a $^3$He density of 4.4 amagats in the target chamber under experimental conditions and a maximum measured $^3$He polarization of 55\%.

\section{\label{sec:level3}Measurement method}

\begin{figure}[ht!]
\centering
\includegraphics[width=3.4in]{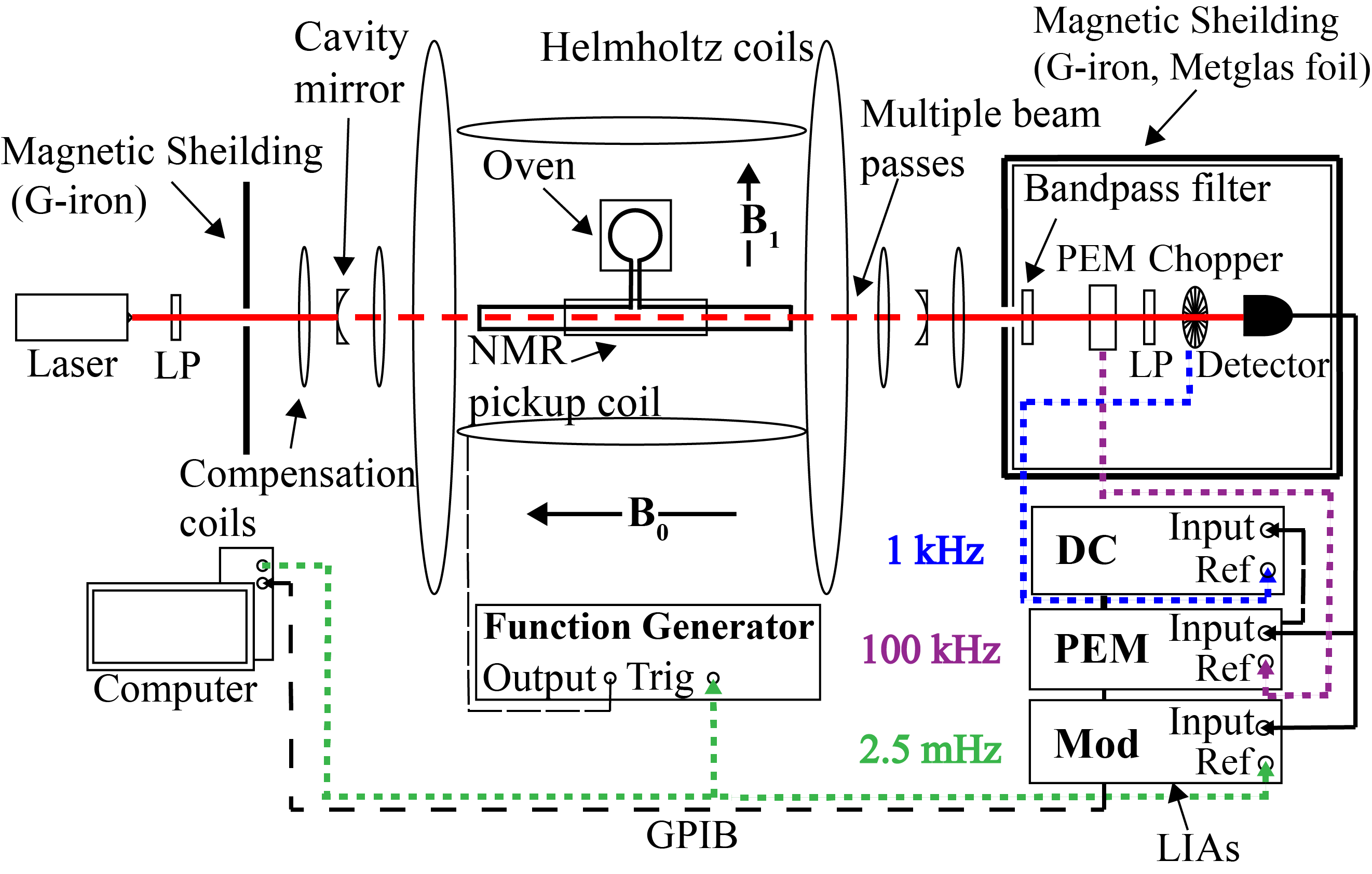}
\caption{Schematic of the experimental setup to measure the gyromagnetic Faraday rotation from polarized $^3$He.}
\label{fig:He_short_paper_1}
\end{figure}

The Faraday rotation is measured with a triple modulation technique in combination with lock-in amplifiers (LIA) \cite{triple_mod_setup}. A schematic of the setup is shown in Fig.(\ref{fig:He_short_paper_1}). Laser light is polarized by a linear polarizer (Glan-Thompson prism), traverses the cavity mirrors and the cell target chamber, then passes into the polarization state analyzing optics encased in magnetic shielding. The optical rotation is converted into a change in intensity by a photoelastic modulator (Hinds I/FS50) and a second linear polarizer oriented at 45$^{\circ}$ to the first polarizer. An optical chopper and a photodetector complete the optical path of the probe beam. The PEM dynamically alters the polarization state of the light at its operating frequency of 50 kHz, detecting at second harmonic gives a signal proportional to the optical rotation. DC drifts are controlled for by measuring the total light intensity at the optical chopper frequency of 1 kHz. To isolate the magnetic Faraday effect of the $^3$He nuclear spins from background static optical rotations, the final modulation is flipping the spins with AFP NMR. The spins switch from parallel to anti-parallel with respect to the static alignment field, each state having a different rotation associated with it. Every flip incurs some polarization loss due to magnetic field gradients, therefore constant spin flipping will result in an equilibrium polarization value smaller than the maximum and given by the balance between the pumping rate and the AFP losses. The optimum rate determined experimentally for the setup is one flip every 200 s yielding a $P_{eq}\sim$17\%. Timing for the spins flips is handled by a computer that controls a function generator powering the transverse RF coils, the computer also sends a 2.5 mHz reference frequency to the final lock-in amplifier. In total there are three lock-in amplifiers that are referenced to the three modulation frequencies, one at 1 kHz for the chopper, one at 100 kHz for the PEM, and one at 2.5 mHz for the spin flipping. Both the DC LIA and the PEM LIA are connected to the detector and the demodulated voltage output of the PEM LIA provides the input for the Spin Mod LIA. The ratio of the voltages from the LIAs combined with a prefactor gives the rotation angle

\begin{equation}
\begin{aligned}
\theta = \frac{G_{PEM}}{12.22}\frac{V_{Mod}}{V_{DC}}.
\end{aligned}
\label{eq:He_short_paper_2}
\end{equation}

The factor G$_{PEM}$ is the sensitivity setting of the PEM LIA in volts and $V_{Mod},V_{DC}$ are the voltages measured by the Spin Mod LIA and DC chopper LIA, respectively. The 12.22 is the result of a combination of various numerical factors needed to extract the true voltage amplitude from the LIAs. To calibrate the system for the small values expected, nanoradian scale Faraday rotation from a glass sample with a known Verdet constant, a material dependent constant that describes the strength of the Faraday rotation, is measured with a square wave magnetic field oscillating at 2.5 mHz. 

To boost the size of the signal expected from $^3$He a multipass cavity is implemented to increase the effective length of the target. Multipass cavities have been previously used to amplify optical rotations as part of sensitive magnetometers \cite{Multipass_magnetometry} and generate large rotation angles for spin noise measurements \cite{Multipass_100rad}. A pair of 2.54 cm spherical concave mirrors with a 1.5 m focal length and coated with a high reflectivity coating form a Herriott style cavity \cite{Herriottcell} with the target chamber in between them. The laser beam reflects back and forth between the mirrors in an elliptical pattern. Each mirror has a 3 mm hole offset 7 mm from the center to allow the laser to enter and exit the cavity. While cylindrical mirrors can achieve a larger number of beam passes \cite{Silver:05}, spherical mirrors were chosen to better focus the beam and compensate for the divergence caused by the end windows on the target chamber. The mirrors sit inside a pair of coils to reduce the ambient longitudinal magnetic field to the 10 mG level. The maximum number of passes for the setup when taking data for the experiment is 13, determined by counting the number of beam spots on the surface of the mirrors. This number is limited by the large distance between the mirrors and intensity losses due to the glass windows on the cell.  

Since the constant flipping of the nuclear spins reduces the equilibrium polarization to about 17\% and the multipass cavity makes the effective length 13 times longer the calculated rotation value from the magnetic Faraday effect changes from 120 nrad to 455 nrad or 35 nrad per pass. The effect does not depend on the frequency of light so this value is true regardless of the probe laser used. Two different lasers were utilized to take measurements, a 633 nm HeNe (Newport R-14309) with a maximum output power of 35 mW and a 405 nm external cavity diode laser (Moglabs LDL 405) with a maximum output power of 60 mW. As a result of larger intensity losses from the windows for blue light the maximum number of passes is reduced to 9, with the expected signal still in the measurable range. 

\begin{figure}[ht!]
\centering
\includegraphics[width=3.4in]{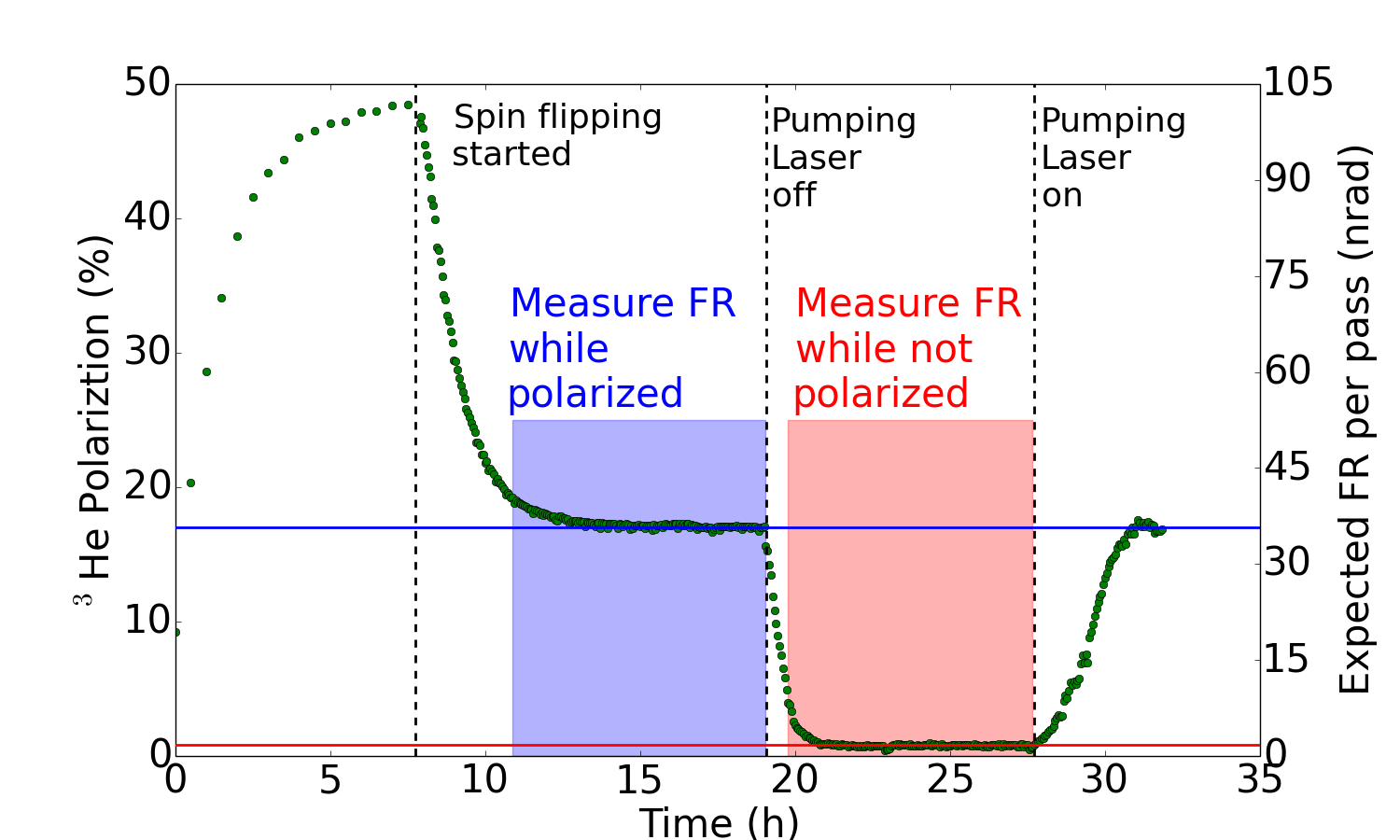}
\caption{Helium nuclear spin polarization during a measurement cycle. The polarization is shown on the left axis and the right shows the predicted magnetic Faraday rotation from the theory at that polarization. Data is collected while the polarization is at equilibrium value and while it is effectively zero.}
\label{fig:He_short_paper_2}
\end{figure}

The magnetic Faraday rotation is measured in the following way, the $^3$He cell is polarized to near maximum along the axis of the holding magnetic field with a magnitude of $B_0 = 21.5$ G. When the polarization is high an EPR calibration measurement is performed and the spin modulation is started via the computer, every 200 s the $B_1 = 200$ mG transverse oscillating magnetic field is turned on to flip the spins through frequency sweep NMR. The polarization is measured during every flip with the NMR pickup coils on the sides of the target chamber. At the same time, the quarter wave plates that convert the light from the pumping laser to circular polarization are rotated so the target continues to be pumped in the correct state. Once the equilibrium polarization is reached the data collection is started for the three lock-in amplifiers. As a result of the low frequency integration time constants of 1000 s are applied to the Spin Mod and DC LIAs. The optical rotation is measured for several hours while the target is polarized to collect enough statistics before the pumping laser is turned off and the $^3$He polarization is allowed to decay to near zero. Without changing any other parameters, the optical rotation is measured while the target is not polarized. Fig.(\ref{fig:He_short_paper_2}) shows the polarization and expected Faraday rotation during a typical measurement. The two data sets can then be compared to see if there is a magnetic Faraday effect from the $^3$He nuclear polarization. After enough statistics have been collected for the non-polarized target the pumping laser can be turned on again and the polarization will increase back to the equilibrium value. The measurement cycle can then be repeated. 

\section{\label{sec:level4}Results}

\begin{figure}[h]
\centering
\includegraphics[width=3.4in]{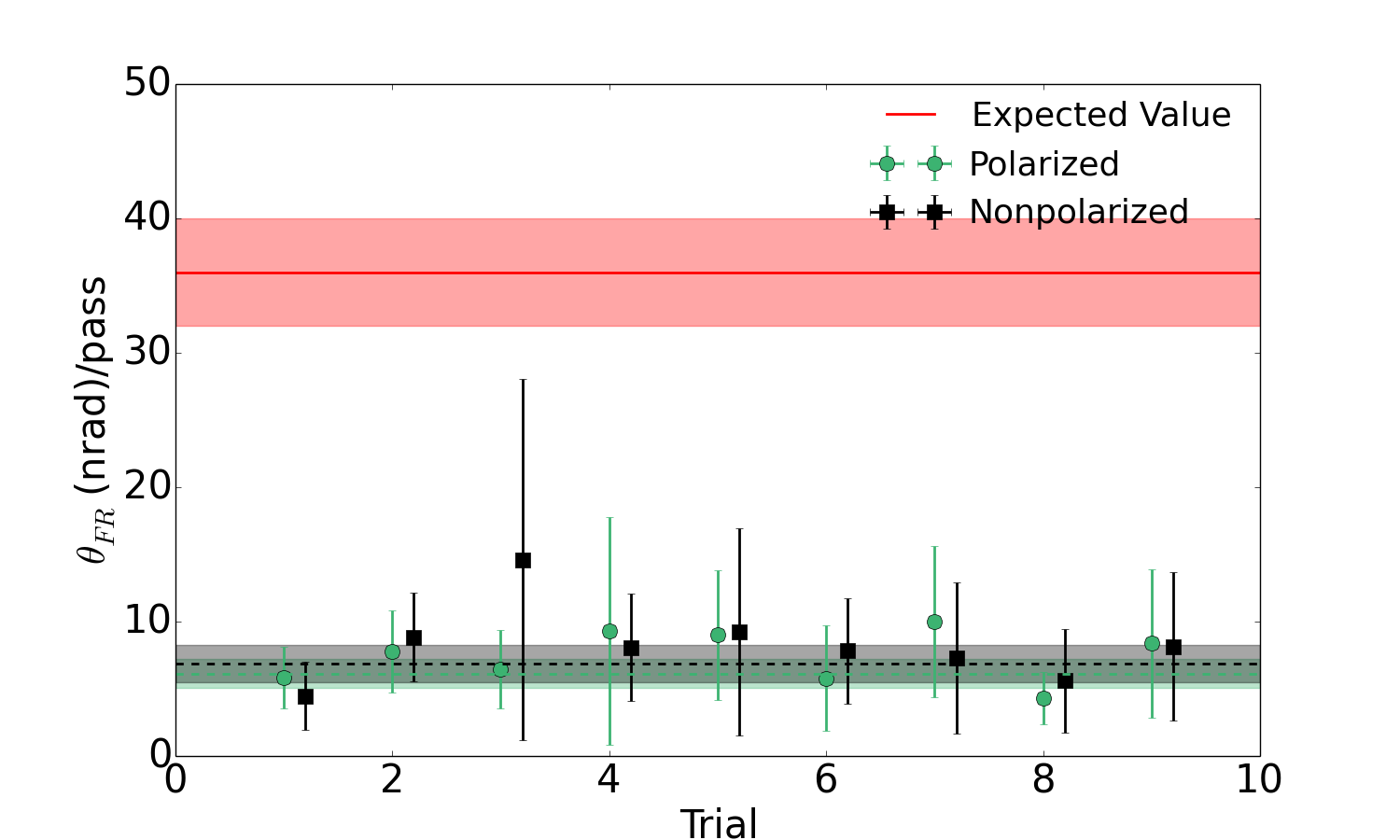}
\caption{Measurements of the gyromagnetic Faraday rotation with a probe laser wavelength of 633 nm and 13 passes in the multipass cavity. Each pair of points corresponds to one measurement cycle of sequential polarized/unpolarized data sets with the averages of all the polarized markers and unpolarized markers given by the dashed lines.}
\label{fig:He_short_paper_3}
\end{figure}

\begin{figure}[h]
\centering
\includegraphics[width=3.4in]{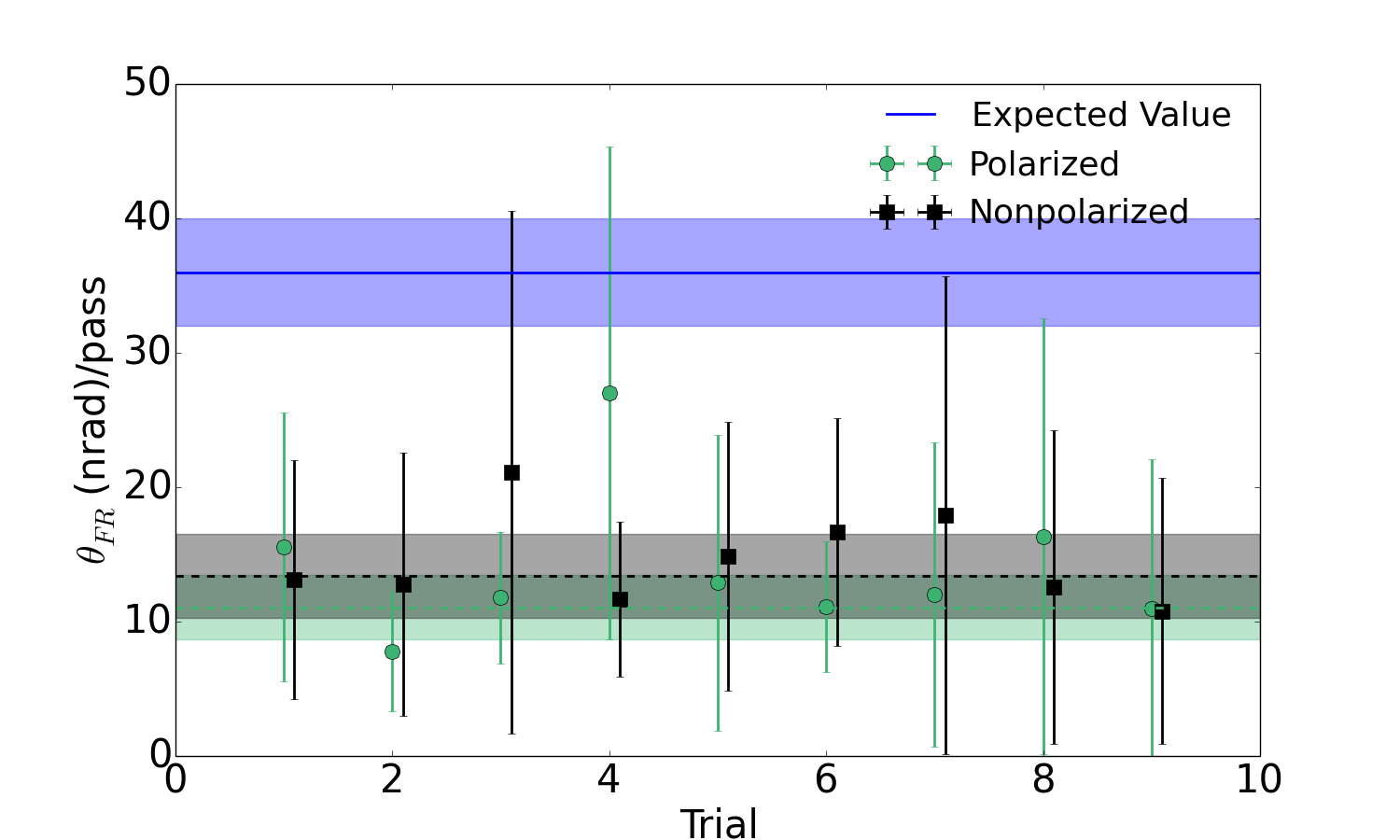}
\caption{Measurements of the gyromagnetic Faraday rotation with a probe laser wavelength of 405 nm and 9 passes in the multipass cavity. For more details see Fig.(\ref{fig:He_short_paper_3}) caption.}
\label{fig:He_short_paper_4}
\end{figure}

Several data sets were taken for the 633 nm HeNe laser and the results are shown in Fig.(\ref{fig:He_short_paper_3}). Each pair of polarized and non-polarized data points that were collected sequentially are plotted next to one another for comparison. A total of nine data sets are shown with an average $^3$He polarization of 16.7$\pm$1.5\%. Furthermore, theoretical Faraday rotation for this polarization is displayed along with the calculated error given by the shaded region. Error bars on the points are statistical only and show one standard deviation. The weighted averages for the polarized target data points and non-polarized target data points are the dashed lines. Not only do the measurements for the polarized target not match the expected value but they exhibit no difference from the non polarized measurements indicating that the effect was not observed. The process was repeated with the 405 nm external cavity diode laser, Fig.(\ref{fig:He_short_paper_4}) shows a similar plot where the average $^3$He polarization was 17.0$\pm$1.5\%. The larger error bars for the 405 nm wavelength are most likely due to more intensity noise in the laser. Again, there appears to be no effect due to the helium nuclear spins and the measured rotations are well below the theoretical prediction. Table \ref{tab:weighted averages} lists the averages for both wavelengths and the theory calculation.        

\begin{table}[h!]
\begin{center}
\caption{Weighted averages of the results for both laser wavelengths.}
\bigskip
\begin{tabular}{ p{1.7cm}p{1.8cm}p{1.8cm}p{1.8cm} }
 \hline
 Wavelength & Polarized\newline Average\newline(nrad/pass) & Non-Polarized Average\newline(nrad/pass) & Theory\newline(nrad/pass)   \\ 
 \thickhline
 633 nm & 6$\pm$1 & 7$\pm$1 & 36$\pm$4  \\
 405 nm & 11$\pm$2 & 14$\pm$3 & 37$\pm$4  \\ 
 \hline
\end{tabular}
\label{tab:weighted averages}
\end{center}
\end{table}

\section{\label{sec:level6}Systematics}

Due to the fact that no signal from magnetically induced Faraday rotation was observed, an array of systematics was investigated. Simple tests such as performing spectral analysis using FFT on the digital output of PEM LIA to detect the modulation signal, rather than using a third LIA, and changing the magnitude of the static magnetic field $|{\mathbf {B_{\mathbf {0}}}}|$ were quick to perform and yielded null results as expected. Those effects most likely to have an impact will be briefly mentioned in this section and a full accounting of the systematics tested and the results are listed in Table \ref{tab:systematics}.

\subsection{Measure with higher polarization}

Before reaching the equilibrium value the $^3$He polarization decays from a higher value after the spin modulation is started (see Fig(\ref{fig:He_short_paper_2})). Though the value is changing it remains larger than $P_{eq}$ for a few hours and the lock-in amplifiers should be able to measure the changing quantity. The average expected Faraday rotation over the range is 1 $\mu$rad which is well above the noise floor. Measurements were taken during this time period as well and the results again show no correlation with theory and are shown in Table. \ref{tab:high pol}

\begin{table}[h]
\begin{center}
\caption{Faraday rotation measured at relatively high $^3$He polarization values. The data were taken during the polarization decay from $P_{max}$ to its equilibrium value. The expected Faraday rotations are larger due to larger $^3$He polarization.}
\label{tab:high pol}
\begin{tabular}{ p{1.7cm}p{1.8cm}p{1.8cm}p{1.8cm} }
 \hline
 Wavelength\newline (nm) & Average polarization (\%) & Polarized (nrad/pass)  & Theory (nrad/pass)   \\ 
 \thickhline
 633 & 36.0 & 3$\pm$1 & 79$\pm$4 \\ 
 & 35.2 & 9$\pm$5 & 77$\pm$4 \\ 
 & 30.1 & 8$\pm$3 & 66$\pm$4 \\ 
 & 29.5 & 7$\pm$5 & 65$\pm$4 \\ 
 \hline
  405 & 34.7 & 23$\pm$13 & 109$\pm$4 \\
 \hline
\end{tabular}
\end{center}
\end{table}

\subsection{Different spin modulation frequencies}

Additional data were taken at a spin flipping frequencies of 5 mHz and 10 mHz The more frequent spin flipping means P$_{eq}$ is at a lower level, averaging at 7.7\% and 3.6\% for 5 mHz and 10mHz, respectively. The expected magnetic Faraday rotation is still expected to be measurable at these polarizations; however, results with the 633 nm probe laser are the similar to the previous spin modulation with no difference between a polarized and unpolarized target.

\subsection{Errors due to laser intensity}

It has been shown previously that lasers with a high intensity focused to a narrow beam width can have a self induced birefringence \cite{Laser_selfbire_1,Laser_selfbire_2,Laser_selfbire_3}. Though the probe lasers seem far away from these extreme effects with typical beam widths of 2 mm, the power was varied from several mW down to a few hundred $\mu$W for both lasers. No change was observed in the rotation measurements other than an increase in the statistical noise.

\subsection{Errors due to pumping laser light}

A false signal at the spin modulation frequency due to a very small amount of pumping laser light leaking into the analyzing optics was discovered. The light was scattered from the pumping chamber and is modulated at the same frequency of the spins due to the rotation of the quarter-wave plates. This was corrected for by adding bandpass filters for the probe lasers along with additional shielding to block the light.

\subsection{Cavity Mirrors}

\begin{figure}[h]
\centering
\includegraphics[width=3.4in]{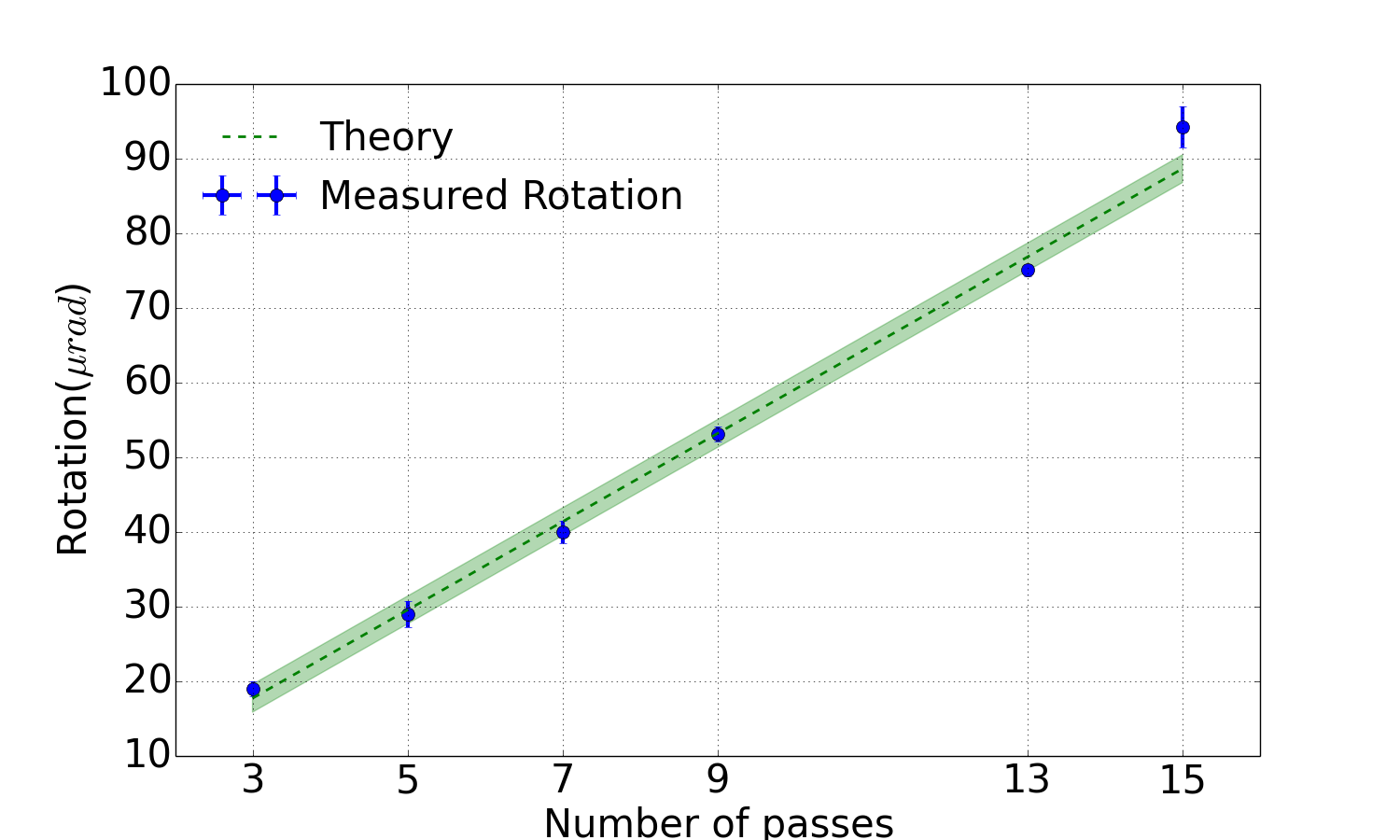}
\caption{Faraday rotation angle from a Corning 1723 glass piece for different number of passes through the cavity. The results are compared to the theory.}
\label{fig:He_short_paper_6}
\end{figure}

To confirm that there were no additional errors from the cavity mirrors the rotation angle from the same glass sample used to calibrate the triple modulation system was measured for different number of passes in the multipass cavity. The results are shown in Fig.(\ref{fig:He_short_paper_6}). The Verdet constant for a 0.32 cm thick Corning 1723 glass piece has been measured precisely ($\mathcal{V} = 3.15\pm0.02$ $\mu$rad/(G cm)) so the results can be compared to the theory and were found to agree. The cavity mirrors seem to have no effect on the rotation angle measurement to the level required for the magnetic Faraday effect experiment.

\subsection{Contribution from glass cell windows and other gases}

There are several corrections to look for due to classical finite magnetic field generated by the polarized $^3$He atoms. This field is modulated at the same frequency as the spins and could also create false signals. The glass windows of the target chamber, Rb vapor, K vapor, and N$_2$ gas are all inside this magnetic field. Similarly, the $^3$He will generate an electric Faraday effect due to its own magnetic field. The Verdet constants for the glass (Corning 1723) and N$_2$ are well known, the calculated Faraday rotation from these expected at this field strength is below the sensitivity of the apparatus. The calculated values are listed as part of Table. \ref{tab:systematics}. To determine what effect the alkali gases might have, the resonant Faraday rotation at the D2 line of Rb has been measured in a 10 cm test cell at room temperature. When the same measurement is performed on the target chamber of the helium cell no optical rotation is measured indicating that the density of Rb vapor is too low to be detected on resonance. Likewise, resonant Faraday rotation from the K 404.721 nm resonance is also measured in a test cell with similar results. Thus, at the far off resonance laser wavelengths used for the gyromagnetic Faraday rotation no contribution from Rb or K is expected. Finally, using the known Verdet constant for the electric part of the Faraday effect \cite{He_Verdet} from $^3$He, the electrically induced rotation is calculated to be too small to be detected at magnetic fields generated by the target magnetization. 

\newcommand{\minitab}[2][c]{\begin{tabular}{#1}#2\end{tabular}}
\begin{table}
\renewcommand{\arraystretch}{1.1}
\begin{center}
    \caption{Summary of systematic studies.} 
    \bigskip
    \begin{tabular}{c|c}
    \hline
    {\bf {Systematic Effect}} & {\bf {Contribution to FR}} \\ 
    \thickhline
    {\minitab[c]{Vary $|{\mathbf {B_{\mathbf {0}}}}|$ between \\ 10 G and 21 G}} & No detectable effect \\ 
    \hline
	{\minitab[c]{Vary probe laser power \\ (28 mW $\to$ 500 $\mu$W) }} & No detectable effect \\ 
	\hline
	{\minitab[c]{Change of DAQ \\ (Use FFT instead of LIA)}} & No detectable effect \\ 
	\hline 
    {\minitab[c]{Change NMR  \\ flipping frequency \\ (2.5 mHz $\to$ 10 mHz)}} & No detectable effect \\ 
    \hline
    {\minitab[c]{Vary target polarization \\ (P$ \approx 55$\% $\to$ P$ \approx 16$\%)}} & {No change}   \\ 
    \hline
    {\minitab[c]{Vary number of passes in \\ Herriott cavity (15 $\to$ 5) }} &  No detectable effect  \\
    \hline
    {\minitab[c]{Higher order theory \\ corrections (in frequency)}} &  $<$1\% of expected rotation \\ 
    \hline
    {\minitab[c]{Correction due to spin \\ misalignment with the \\ probe laser}} &  $<$1\% of expected rotation  \\
    \hline
    {\minitab[c]{Contribution from electric \\ Faraday effect from $^3$He}}  &  {\minitab[c]{0.019 nrad/pass (633 nm) \\   0.065 nrad/pass (405 nm)}}  \\ 
    \hline
    {\minitab[c]{Contribution from cell \\ end windows}} &  {\minitab[c]{1 nrad/pass (633 nm) \\ 2 nrad/pass (405 nm)}} \\ 
    \hline
    {\minitab[c]{Errors due to cavity \\ mirrors}} & {\minitab[c]{Signals scale as expected, \\ no  detectable effect \\ observed}}  \\ 
    \hline
    {\minitab[c]{Depolarization due to \\ cell window misalignment }}  &  {{$\ll 0.5$\%}} \\ 
    \hline
    {\minitab[c]{False signal from \\ pumping laser light }}  & {\minitab[c]{Corrected with bandpass \\ filters}} \\ 
    \hline
    {\minitab[c]{Faraday effect from Rb in \\ the target chamber due to $^3$He \\ magnetic field}} & No detectable effect  \\ 
    \hline 
    {\minitab[c]{Calculated Faraday effect from N$_2$ \\ in the target chamber due to $^3$He \\ magnetic field }} & {\minitab[c]{0.008 nrad/pass (633 nm) \\ 0.015 nrad/pass (405 nm)}} \\ 
    \hline
    {\minitab[c]{Calculated Faraday effect from air \\ outside windows due to $^3$He \\ magnetic field}} & {\minitab[c]{0.003 nrad/pass (633 nm) \\  0.007 nrad/pass (405 nm) }}  \\ 
    \hline
    \end{tabular}
    \label{tab:systematics}
\end{center}
\end{table}

\section{\label{sec:level7}Conclusions}

In summary, the gyromagnetic Faraday effect is a predicted optical rotation induced by magnetic moments. It is a purely magnetic effect that is separate from the standard Faraday effect which results from the coupling of light to electric polarizabilities or atomic or molecular resonances. Theoretically, the rotation angles expected are significantly smaller and exhibit no frequency dependence on the probe light which would allow the effect to be isolated from other optical rotations. We used a dense hyperpolarized $^3$He gas as a test system to attempt to discover the effect because it has many favorable qualities that lend to isolating the signal. A $^3$He target was polarized with spin exchange optical pumping and combined with a sensitive apparatus that can detect optical rotations several times smaller than the expected signal size. Several measurements at two different probe laser wavelengths were collected and compared with the expected value. The results showed no nuclear spin induced rotation in the predicted range. Additionally, we observed no difference between a polarized and an unpolarized target indicating that the effect is even smaller than the limit that the setup can measure with an upper limit of $7\pm1$ nrad/pass given by the weighted average of the polarized measurements. One difference that should be noted is that the theory and estimates were for bare magnetic moments while the experiment was performed with entire atoms. The interaction of light with more complex systems might alter the coherent accumulation of the Faraday rotation angle \cite{FR_beyond_theory}.

\begin{acknowledgments}
This work is partially supported by the U.S. Department of Energy Office of Nuclear Physics under contract No. DE-FG02-99ER41101 and a research grant awarded to W.K. by the University of Kentucky. The authors would like to thank Todd Averett for filling the helium-3 cell and the determination of the gas density and they would like to thank Susan Gardner for useful discussions.  W.K. would like to express special thanks to the Mainz Institute for Theoretical Physics (MITP) for its hospitality and support. 
\end{acknowledgments}
\bibliography{main}
\end{document}